\documentclass[twocolumn,aps,pre,longbibliography]{revtex4-1}
\usepackage{amsmath,amssymb}
\usepackage{graphicx}
\usepackage{dcolumn}
\usepackage{color}
\usepackage{epstopdf}
\usepackage[%
colorlinks=true,
urlcolor=blue,
linkcolor=blue,
citecolor=blue
]{hyperref}
\begin{document}
\title{Steepest-entropy-ascent quantum thermodynamic modeling of the far-from-equilibrium interactions between nonequilibrium systems of indistinguishable particle ensembles}
\author{Guanchen Li}
\email{guanchen@vt.edu}
\author{Michael R. von Spakovsky}
\email{vonspako@vt.edu}
\affiliation{%
	Center for Energy Systems Research, Mechanical Engineering Department\\
	Virginia Tech, Blacksburg, VA 24061
}%
\date{\today}

\begin{abstract}
	This paper presents a nonequilibrium, first-principles, thermodynamic-ensemble based model for the relaxation process of interacting non-equilibrium systems. This model is formulated using steepest-entropy-ascent quantum thermodynamics (SEAQT) and its equation of motion for a grand canonical ensemble and is applied to a many particle system of classical or indistinguishable particles. Two kinds of interactions are discussed, including pure heat diffusion and heat and mass diffusion together. Since no local equilibrium assumption is made, the conjugate fluxes and forces are intrinsic to the subspaces of the state space of one system and/or of the state space of the two interacting systems. They are derived via the concepts of hypoequilibrium state and nonequilibrium intensive properties, which describe the nonmutual equilibrium status between subspaces of the thermodynamic state space of a single system and/or of the state space of the two interacting systems. The Onsager relations are shown to be thermodynamic kinematic features of the system and are found without knowledge of the detailed mechanics of the dynamic process. A fundamental thermodynamic explanation for the measurement of each intensive property of a system in a nonequilibrium state is given. The fundamental thermodynamic definition of reservoir is also discussed. Finally, the equation of motion for a system undergoing multiple interactions is provided, which permits the modeling of a network of local systems in nonequilibrium at any spatial and temporal scale.
\end{abstract}
\maketitle
\section{Introduction}
The study of nonequilibrium relaxation processes - including chemical kinetics, mass diffusion, and heat diffusion – is typically accomplished using approaches based on microscopic mechanics \cite{Rapaport2004,Chen1998,Vogl2010,Newman1999} or thermodynamics \cite{Groot1962,Gyarmati1970,Kubo2012,Seifert2012,Jou1996}. Approaches based on thermodynamics are able to generally capture the features of the relaxation process via, for example, the Onsager relations. The thermodynamic features captured can be regarded as a coarse graining of the microscopic dynamics or as a pattern in ensemble evolution \cite{Grmela1997,Ottinger1997}, which computationally is more efficient. However, most of these approaches have limited or no applicability in the far-from-equilibrium realm, since the local or near-equilibrium assumption is needed or only analytical solutions at steady state are available. In addition, their governing equations are phenomenological or stochastic in nature and, thus, do not have a first-principles basis. To address these issues and push the application of thermodynamic principles further into the nonequilibrium realm, it is of great importance to find a general and simple description of nonequilibrium state corresponding to a thermodynamic pattern of the microscopic description, to fundamentally define the macroscopic properties of any thermodynamic state (i.e., extensive or intensive properties for both equilibrium and nonequilibrium states), and to use a thermodynamic governing  equation based on first principles.

Steepest-entropy-ascent quantum thermodynamics (SEAQT), which is a first-principles, thermodynamic-ensemble based approach, addresses all of the issues raised above, providing a governing equation able to describe the nonequilibrium process from an entropy generation viewpoint. The macroscopic properties of entropy, energy, and particle number, which are well defined for any state of any system \cite{Zanchini2014}, are used to develop the governing equation and describe system state evolution. Recently, this description has been further simplified via the concept of hypoequilibrium state \cite{LivonSpakovsky2015,LivonSpakovsky2015a}, which captures the global features of the microscopic description for the relaxation process. In addition, the concept of nonequilibrium intensive properties introduced in \cite{LivonSpakovsky2015,LivonSpakovsky2015a} based on the concept of hypoequilibrium state enables a complete description of the nonequilibrium evolution of state when combined with the set of nonequilibrium extensive properties. Unlike the intensive property definitions of other nonequilibrium thermodynamic approaches (definitions which require the local-equilibrium, near-equilibrium, or steady state assumption or a phenomenological basis), the definitions in the SEAQT framework are fundamental and available to all nonequilibrium states and are especially suitable for the description of the evolution in state of relaxation processes. Both of these concepts enable the generalization of many equilibrium (or near-equilibrium) thermodynamic relations such as the Gibbs relation, the Clausius inequality, and the Onsager relations into the far-from-equilibrium realm as well as for non-quasi-equilibrium processes.

In this paper, SEAQT is applied to the study of the interaction of systems using the grand partition function. The system studied can be any distinguishable or indistinguishable system with or without long distance intermolecular interaction. The evolution of two systems can be a non-quasi-equilibrium process. In Sec. II, the equation of motion and the concepts of hypoequilibrium state and nonequilibrium intensive properties are introduced. In Sec. III, interacting systems with heat diffusion only are studied. The Onsager relations and a thermodynamic explanation of measurement (of a system in equilibrium or nonequilibrium) and reservoir is given. In Sec. IV, interacting systems with heat and mass diffusion are studied followed in Sec. V by the study of a system interacting with multiple systems and a discussion of the applicability of the SEAQT framework to the network of nonequilibrium systems.

\section{SEAQT equation of motion}
\subsection{General equation of motion}
In this section, the system and state description in SEAQT is given, and the equation of motion, is presented. Based on the discussion by Grmela \cite{Grmela2014,Grmela1997,Ottinger1997} and Beretta \cite{Beretta2014,Montefusco2015} the general form of a nonequilibrium framework is a combination of both irreversible relaxation and reversible symplectic dynamics. If written in a generalized form of Ginzburg-Landau equation \cite{Grmela1997,Montefusco2015}, the equation of motion takes the following form:
\begin{equation}
\frac{d}{dt}\phi(t)=X^{H}_{\phi(t)}+Y^{H}_{\phi(t)}
\end{equation}
where $\phi(t)$ represents the state evolution trajectory, $X^{H}_{\phi(t)}$ and $Y^{H}_{\phi(t)}$ are functions of the system state $\phi(t)$ and represent the reversible symplectic and irreversible relaxation dynamics, respectively. In the SEAQT framework, the system is defined by the Hamiltonian operator $\hat{H}$, the system state is represented by the density operator $\hat{\rho}$, $X^{H}_{\phi(t)}$ follows the Schr\"{o}dinger equation, and $Y^{H}_{\phi(t)}$ is derived from the SEA principle. To describe the evolutionary process, conservation laws are explicitly required in order to construct the equation of motion, which is given by \cite{Beretta2009}
\begin{equation}
\frac{d\hat{\rho}}{dt}=\frac{1}{i\hbar}[\hat{\rho},\hat{H}]+\frac{1}{\tau}\hat{D}
\end{equation}
where the first term is the Schr\"{o}dinger term, and the second is the dissipation term. If the system is in pure (zero-entropy) state, $\hat{\rho}\hat{\rho}=\hat{\rho}$, and, the equation of motion reverts back to the Schr\"{o}dinger equation of quantum mechanics. If the system is in a so-called mixed (nonzero-entropy) state and $\hat{\rho}$ is diagonal in the energy eigenstate basis, $\hat{H}$ commutes with $\hat{\rho}$ and the Schr\"{o}dinger term goes to zero even though $\hat{\rho}$ may not be a Maxwellian distribution among the energy eigenlevels. The state evolution of such a mixed-state operator cannot be captured by the Schr\"{o}dinger term and is instead given by the second term to the right of the equals, the dissipation term, which captures the probability redistribution towards the Maxwellian distribution. The dissipation term is constructed using a set of operators called the `generators of the motion'. Each generator corresponds to one of the conservation laws to which the system is subjected. For example, an nonreacting isolated system is subject to two conservation laws, probability normalization and energy conservation, so that the generators of the motion are $\{\hat{I},\hat{H}\}$. 

In the study of two interacting systems, the system state space is given by
\begin{equation}
\mathcal{H}=\mathcal{H}_a\otimes\mathcal{H}_b
\end{equation}
where $\mathcal{H}_{a(b)}$ are the Hilbert space of two systems $a(b)$ (for a general system with a variable number of indistinguishable particle, $\mathcal{H}_{a(b)}$ will be Fock space) and the initial state of density operator is chosen to be  
\begin{equation}
\hat{\rho}=\hat{\rho}_a\otimes\hat{\rho}_b
\end{equation}
Note that it does not include a correlation term. The Hamiltonian operator of the system is then
\begin{eqnarray}
\hat{H}=\hat{H}_a\otimes\hat{I}_b+\hat{I}_a\otimes\hat{H}_b
\end{eqnarray}
where no interparticle interaction term has been included. If we assume both system $a$ and system $b$ to consist of a dilute-Boltzmann gas that give the diagonal density operators $\hat{\rho}_a$ and $\hat{\rho}_b$, the equation of motion reduces to
\begin{equation}
\frac{dp_i^{a(b)}}{dt}=\frac{1}{\tau}D_i^{a(b)}(\boldsymbol{p})
\end{equation}
where $p_i^a(b)$ is the $i$th diagonal term of $\hat{\rho}^{a(b)}$ in the energy eigenstates basis and represents the probability of the system $a(b)$ being in the eigenstate associated with the $i$th energy eigenlevel, $\boldsymbol{p}$ represents the distributions $\{p_i^a,\,i=1,\cdots\}$ and $\{p_j^b\,j=1,\cdots\}$ for system $a$ and $b$, and $\tau$ is the relaxation time.
\subsection{Nonequilibrium state and state evolution description: Hypoequilibrium}
The thermodynamic features of the nonequilibrium relaxation process generated by the SEAQT framework have a number of useful characteristics, which lead to a complete description of nonequilibrium state and the general fundamental definition of nonequilibrium intensive properties. This description is based on two key concepts, hypoequilibrium state and nonequilibrium intensive properties, which are briefly discussed below.

For a given system, such as system $a$, represented by an energy eigenlevel set $\Omega^a = \{(n_i^a,\epsilon_i^a,N_i^a)\}$, where each energy eigenlevel is represented by a triplet of energy ($\epsilon_i^a$) and particle number ($N_i^a$) eigenvalues and by its degeneracy ($n_i^a$), the system can be divided into $M_a$ subsystems $\Omega_K^a = \{(n_i^{a,K},\epsilon_i^{a,K},N_i^{a,K})\}, K=1,\cdots,M_a$, so that the state space of system $a$ (Hilbert space) $\mathcal{H}^a$ can be represented by the sum of $M_a$ subspaces $\mathcal{H}_K^a$ with $K=1,2,...,M_a$.
\begin{equation}
\mathcal{H}^a=\bigoplus_{K=1}^{M_a}\mathcal{H}_K^a
\end{equation}
To be complete, $M_a$ can be infinite. The state of system $a$ can be represented by the distributions in $M_a$ subspace energy eigenlevels $\{p_i^{a,K}, K=1,...,M_a\}$.

If the probability distribution in each subsystem yields to a grand canonical distribution, the system is designated as being in an $M_a$th-order hypoequilibrium state. Based on this definition, it can be shown that any state of the system $a$ is a hypoequilibrium state with order $M_a$, where $M_a$ is less than or equal to the number of system eigenlevels \cite{LivonSpakovsky2015,LivonSpakovsky2015a}. A hypoequilibrium state of order 1 corresponds to a state in stable equilibrium. The probability distribution of the $K$th subspace of the $M_a$th-order hypoequilibrium state takes the form
\begin{equation} p_i^{a,K}=\frac{p^{a,K}}{\Xi^{a,K}(\beta^{a,K},\gamma^{a,K})}e^{-\beta^{a,K}\epsilon_i^{a,K}-\gamma^{a,K}N_i^{a,K}}
\end{equation}
where $\beta^{a,K}$ and $\gamma^{a,K}$ are parameters, $p^{a,K}$ is the total probability in subspace $K$ of system $a$, and $\Xi^{a,K}(\beta^{a,K},\gamma^{a,K})$ is the grand partition function of the subspace with parameters $\beta^{a,K}$ and $\gamma^{a,K}$. To be complete, $\beta^{a,K}=0$ and $\gamma^{a,K}=0$ if $\#(\mathcal{H}_a^K)=1$, $\gamma^{a,K}=0$ if $\#(\mathcal{H}_a^K)=2$ and the $\#(\mathcal{H}_a^K)$ can be infinite. The grand partition function is written as
\begin{equation}
\Xi^{a,K}(\beta^{a,K},\gamma^{a,K})=\sum_{K=1}^{\#(\mathcal{H}_a^K)} n_i^{a,K}e^{-\beta^{a,K}\epsilon_i^{a,K}-\gamma^{a,K}N_i^{a,K}}
\end{equation}
Then, defining
\begin{equation}
\alpha^{a,K} = \ln \Xi^{a,K}(\beta^{a,K},\gamma^{a,K})-\ln p^{a,K}
\end{equation}
so that the probability distribution of the Kth subspace can be represented using $\{(\alpha^{a,K},\beta^{a,K},\gamma^{a,K}),\, K=1,\cdots,M_a\}$, i.e.,
\begin{equation}
p_i^{a,K} = n_i^{a,K}e^{-\alpha^{a,K}}e^{-\epsilon_i^{a,K}\beta^{a,K}}e^{-N_i^{a,K}\gamma^{a,K}}
\end{equation}

For a given $M_a$th-order hypoequilibrium state, the intensive properties of the subspaces can be represented by $\beta^{a,K}$ and $\gamma^{a,K}$ or equivalently using temperature and chemical potential defined by
\begin{equation}
T^{a,K}=\frac{1}{k_b\beta^{a,K}},\quad\mu^{a,K}=\gamma^{a,K}T^{a,K}
\end{equation}
A $M_a$th-order hypoequilibrium state can then be represented by a division $\{\Omega_K = (n_i^{a,K},\epsilon_i^{a,K},N_i^{a,K}),\ K=1,...,M_a\}$ of the system and a corresponding triplet set $\{(\alpha^{a,K},\beta^{a,K},\gamma^{a,K}),\ k=1,...,M_a\}$. The intensive property set $\{(T^{a,K},\mu^{a,K}),\ K=1,...,M_a\}$ is a generalization of the definition of intensive property at stable equilibrium $(T^{eq},\mu^{eq})$, which itself is a first-order hypoequilibrium state. Appendix A proves that for the equation of motions used in this paper, if a system begins in a $M_a$th-order hypoequilibrium state, it remains in a $M_a$th-order hypoequilibrium state throughout the state evolution as long as the same subsystem division is maintained. Thus, the time evolution of the distribution is given by
\begin{eqnarray}
 p_i^{a,K}(t)=n_i^{a,K}e^{-\alpha^{a,K}(t)-\beta^{a,K}(t)\epsilon_i^{a,K}-\gamma^{a,K}(t)N_i^{a,K}}
\end{eqnarray}
The intensive property set $\{(T^{a,K}(t),\mu^{a,K}(t)),\ i=1,...,M\}$ is well defined throughout the entire state evolution. This system state evolution can also be represented by the evolution of the triplet set $\{(\alpha^{a,K}(t),\beta^{a,K}(t),\gamma^{a,K}(t)),\ K=1,...,M\}$. In the discussions below, the triplet $(\alpha^{a,K},\beta^{a,K},\gamma^{a,K})$ are also called intensive properties, since they are equivalent to temperature and chemical potential.

\section{Interacting systems with heat diffusion only}
In this section, system $a$ and system $b$ form a composite system, and only heat diffusion is allowed between them. Both system $a$ and system $b$ can be in nonequilibrium states, and are represented by the probability distribution among the energy eigenlevels of system $a$ and $b$, i.e., by $\{p_i^a\}$ and $\{p_i^b\}$.

\subsection{Equation of motion}
For the case when only heat diffusion is present, five conservation laws hold: probability and particle number conservations for both system $a$ and system $b$, and total energy conservation of the composite system. Thus, the generators of motion are $\{\hat{I}_a,\hat{I}_b,\hat{N}_a,\hat{N}_b,\hat{H}\}$ constrained by 
\begin{eqnarray}
I^a&=&\sum_i p_i^a=1\\
I^b&=&\sum_i p_i^b=1\\
N^a&=&\sum_i N_i^ap_i^a=\text{constant}\\
N^b&=&\sum_i N_i^bp_i^b=\text{constant}\\
E&=&\sum_i \epsilon_i^ap_i^a+\sum_i \epsilon_i^bp_i^b=\text{constant}
\end{eqnarray}
Based on the derivation in Appendix B, the equation of motion for system $a$ takes the form
\begin{eqnarray}\label{EOM_heat}
	\frac{dp_j^a}{dt}=\frac{1}{\tau}\frac{\left|\begin{array}{cccccc}
			-p_j^a\ln \frac{p_j^a}{n_j^a} & p_j^a  & N_j^ap_j^a & 0& 0 & \epsilon_j^ap_j^a \\
			\langle s\rangle^a & 1 & \langle N\rangle^a & 0 & 0 &\langle e\rangle^a \\
			\langle Ns\rangle^a & \langle N\rangle^a  & \langle N^2\rangle^a & 0 & 0& \langle eN\rangle^a\\
			\langle s\rangle^b & 0  & 0 & 1 &\langle N\rangle^b &\langle e\rangle^b \\		
			\langle Ns\rangle^b & 0  & 0 & \langle N\rangle^b & \langle N^2\rangle^b & \langle eN\rangle^b\\
			\langle es\rangle & \langle e\rangle^a  & \langle eN\rangle^a & \langle e\rangle^b& \langle eN\rangle^b & \langle e^2\rangle
		\end{array}\right|}{\left|\begin{array}{ccccc}
		1  & \langle N\rangle^a & 0 & 0 &\langle e\rangle^a \\
		\langle N\rangle^a &  \langle N^2\rangle^a & 0 & 0& \langle eN\rangle^a\\
		0 & 0 & 1 &\langle N\rangle^b &\langle e\rangle^b \\	
		0 & 0 & \langle N\rangle^b & \langle N^2\rangle^b & \langle eN\rangle^b\\
		\langle e\rangle^a & \langle eN\rangle^a &\langle e\rangle^b &  \langle eN\rangle^b & \langle e^2\rangle
	\end{array}\right|}\nonumber\\
\end{eqnarray}
The numerator of the ratio of determinants on the right can be expanded to yield
\begin{equation}
	\det = -p_j\ln \frac{p_j^a}{n_j^a}|C_1|-p_j^a|C_2^a|+N_j^ap_j^a|C_3^a|-\epsilon_j^ap_j^a|C_4|
\end{equation}
where $|C_1|$, $|C_2^a|$, $|C_3^a|$, and $|C_4|$ are the minors of the first line of the determinant. By defining 
\begin{equation}
	\frac{|C_2^a|}{|C_1|}=\alpha_a^0,  \quad\frac{|C_3^a|}{|C_1|}=-\gamma_a^0,
	\quad\frac{|C_4|}{|C_1|}=\beta^0,
\end{equation}
the equation of motion transforms to
\begin{equation}
	\frac{dp_j^a}{dt}=\frac{1}{\tau}(-p_j^a\ln \frac{p_j^a}{n_j^a}-p_j^a\alpha_a^0-N_j^ap_j^a\gamma_a^0-\epsilon_j^ap_j^a\beta^0)
\end{equation}
Now, defining a row vector of extensive properties
\begin{equation}\label{row}
	\vec{l}_j^a=\left[\begin{array}{ccc}
		1 & N_j^a& \epsilon_j^a 
	\end{array}\right]
\end{equation}
where the subscript $j$ refers to the $j$th energy eigenlevel and defining a column vector of intensive properties
\begin{equation}
	\vec{\mu}_a^0=\left[\begin{array}{c}
		\alpha_a^0\\ \gamma_a^0\\ \beta^0
	\end{array}\right]
\end{equation}
the equation of motion can be written as
\begin{equation}\label{EOM_vec}
	\frac{dp_j^a}{dt}=\frac{1}{\tau}(-p_j^a\ln \frac{p_j^a}{n_j^a}-p_j^a\vec{l}_j^a\cdot\vec{\mu}_a^0)
\end{equation}
In stable equilibrium, each element of $\vec{\mu}_a^0$ turns out to be an intensive property of the composite system.

\subsection{Hypoequilibrium state and nonequilibrium intensive properties}
We assume that the initial state of system $a$ is a $M_a$th-order hypoequilibrium state, and that the probability for the $i$th energy eigenlevel represented by the triplet $(n_i^{a,K}, \epsilon_i^{a,K}, N_i^{a,K})$, where $i=1,2,\cdots,\,K=1,2,\cdots M_a$, is given by 
\begin{equation}
p_i^{a,K} = n_i^{a,K}e^{-\alpha^{a,K}}e^{-\epsilon_i^{a,K}\beta^{a,K}}e^{-N_i^{a,K}\gamma^{a,K}}
\end{equation}
where the triplet $\{(\alpha^{a,K},\beta^{a,K},\ \gamma^{a,K}),\, K=1,\cdots,M_a\}$ has been used in the representation. This can be rewritten in terms of row vector of extensive properties and column vector of intensive properties such that
\begin{equation}
\ln \frac{p_i^{a,K}}{n_i^{a,K}} = -\alpha^{a,K}-\epsilon_i^{a,K}\beta^{a,K}-N_i^{a,K}\gamma^{a,K}= -\vec{l}_j^{a,K}\cdot\vec{\mu}^{a,K}
\end{equation}
where the superscripts refer to system $a$ and the $K$th subspace of system $a$, and the subscript to the $i$th energy eigenlevel in the $K$th subspace. The row and column vectors are defined as
\begin{eqnarray}
\vec{l}_i^{a,K}&=&\left[\begin{array}{ccc}
1 & N_i^{a,K}& \epsilon_i^{a,K} 
\end{array}\right]\\
\vec{\mu}^{a,K}&=&\left[\begin{array}{c}
\alpha^{a,K}\\ \beta^{a,K}\\ \gamma^{a,K}
\end{array}\right]
\end{eqnarray}
In Appendix A, it is proven that if the initial state is a hypoequilibrium state, the system remains in a hypoequilibrium state with the same subsystem division, which means that the time evolution of the distribution of system $a$ takes the form
\begin{eqnarray}
p_i^{a,K}(t) &=& n_i^{a,K}e^{-\alpha^{a,K}(t)}e^{-\epsilon_i^{a,K}\beta^{a,K}(t)}e^{-N_i^{a,K}\gamma^{a,K}(t)}\nonumber\\
&=&n_i^{a,K}e^{-\vec{l}_i^{a,K}\cdot\vec{\mu}^{a,K}(t)}
\end{eqnarray}
where the evolution of the intensive properties $\alpha^{a,K}(t),\ \beta^{a,K}(t),\ \gamma^{a,K}(t)$ are the solutions to (see Appendix A)
\begin{eqnarray}
\frac{d\vec{\mu}^{a,K}(t)}{dt}=-\frac{1}{\tau}(\vec{\mu}^{a,K}(t)-\vec{\mu}_a^0(t))
\end{eqnarray}
and the equation of motion Eq. (\ref{EOM_vec}), is expressed as
\begin{equation}
\frac{dp_i^{a,K}}{dt}=\frac{1}{\tau}p_i^{a,K}\vec{l}_i^{a,K}\cdot(\vec{\mu}^{a,K}-\vec{\mu}_a^0)
\end{equation}
Thus, if the initial state of system $a$ is a $M_a$th-order hypoequilibrium and that of system $b$ is a $M_b$th-order hypoequilibrium, only $3(M_a+M_b)$ ODEs need to be solved in order to determine the nonequilibrium evolution.

\subsection{Time evolution of subsystem extensive property and Onsager relations}
Based on the row vector for extensive properties of one energy eigenlevel, the vector of the extensive properties in the $K$th subsystem of system $a$ can be defined as a row vector as well such that
\begin{equation}
\vec{L}^{a,K}=\sum_i p_i^{a,K}\vec{l}_i^{a,K}=\left[\begin{array}{ccc}
p^{a,K} & E^{a,K} & N^{a,K}
\end{array}\right]
\end{equation}
where $p^{a,K}$, $E^{a,K}$, and $N^{a,K}$ are the contributions of the $K$th subsystem to the total extensive properties of system $a$ and are defined by
\begin{eqnarray}
p^{a,K}&=&\sum p_i^{a,K}\\
E^{a,K}&=&\langle e\rangle^{a,K}=\sum \epsilon_i^{a,K}p_i^{a,K}\\
N^{a,K}&=&\langle N\rangle^{a,K}=\sum N_i^{a,K}p_i^{a,K}
\end{eqnarray}

The evolutions of these extensive properties and others are governed by
\begin{eqnarray}\label{37}
\frac{d\vec{L}^{a,K}(t)}{dt}=-\frac{1}{\tau}(\vec{\mu}^{a,K}(t)-\vec{\mu}_a^0(t))^T[C_1]^{a,K}
\end{eqnarray}
where
\begin{equation}
[C_1]^{a,K}=\left[\begin{array}{ccc}
p^{a,K} & \langle e\rangle^{a,K} & \langle N\rangle^{a,K}\\
\langle e\rangle^{a,K} & \langle e^2\rangle^{a,K}  & \langle eN\rangle^{a,K}\\
\langle N\rangle^{a,K} & \langle eN\rangle^{a,K} & \langle N^2\rangle^{a,K}
\end{array}\right]
\end{equation}
Here $\langle\dots\rangle^{a,K}$ is the contribution of the $K$th subspace to the total extensive property $\langle\dots\rangle^{a}$ of system $a$. 

The rate of entropy change of the $K$th subsystem is then
\begin{eqnarray}
&&\frac{dS^{a,K}}{dt} = \frac{d\langle s\rangle^{a,K}}{dt}=\sum_i\frac{d}{dt}(-p_i^{a,K}\ln \frac{p_i^{a,K}}{n_i^{a,K}})\nonumber\\
&&=\sum_j(-\ln \frac{p_i^{a,K}}{n_i^{a,K}}-1)\frac{dp_i^{a,K}}{dt}=\sum_j(\vec{l}_i^{a,K}\cdot\vec{\mu}^{a,K}-1)\frac{dp_i^{a,K}}{dt}\nonumber\\
&&=\frac{d\vec{L}^{a,K}}{dt}\cdot\vec{\mu}^{a,K}-\frac{dp^{a,K}}{dt}
\end{eqnarray}
and for system $a$
\begin{eqnarray}
\frac{dS^a}{dt} &=& \sum_K\frac{dS^{a,K}}{dt} =\sum_K\frac{d\vec{L}^{a,K}}{dt}\cdot\vec{\mu^{a,K}}
\end{eqnarray}
where probability conservation for system $a$ has been used. The rate of system entropy is then for the composite system is
\begin{eqnarray}\label{41}
\frac{dS}{dt}&=&\sum_a\frac{dS^a}{dt}=\sum_a\sum_K\frac{d\vec{L^{a,K}}}{dt}\cdot\vec{\mu^{a,K}}\nonumber\\
&=&\sum_a\sum_K\frac{d\vec{L^{a,K}}}{dt}\cdot(\vec{\mu}^{a,K}-\vec{\mu_a}^0)
\end{eqnarray}
where in the last equal sign, the conservation laws has been applied. Defining conjugate fluxes and conjugate forces, respectively, as
\begin{eqnarray}
\vec{J}^{a,K}=\frac{d\vec{L}^{a,K}}{dt},\,\vec{X}^{a,K}=\vec{\mu}^{a,K}-\vec{\mu}_a^0
\end{eqnarray}
the evolutions of the extensive properties can be written from Eqs. (\ref{37}) and (\ref{41}) as
\begin{eqnarray}
\vec{J}^{a,K}&=&\vec{X}^{a,K}\cdot[C_1]^{a,K}\\
\frac{dS}{dt}&=&\sum_a\sum_K\vec{J}^{a,K}\vec{X}^{a,K}
\end{eqnarray}
Since $[C_1]^{a,K}$ is positive definite and symmetric, the Onsager relations are acquired. Specifically, due to the conservation laws, probability and particle number fluxes occur within a given system $a$ or $b$, while the energy fluxes can cross from one system to the next. Thus, the Onsager relations hold here for the relaxation process of system $a$ and for the non-quasi-equilibrium process between systems $a$ and $b$.

\subsection{Linkage between two systems: Measurement and reservior}
The conservation of probability and particle number for system $a$ leads to
\begin{eqnarray}
\frac{dp^a}{dt}&=&\sum_K\frac{dp^{a,K}}{dt}=0\\
\frac{d\langle N\rangle^a}{dt}&=&\sum_K\frac{d\langle N\rangle^{a,K}}{dt}=0
\end{eqnarray}
which can be written as 
\begin{eqnarray}
&&\sum_K(\alpha^{a,K}-\alpha_a^0)p^{a,K}+\sum_K\langle e\rangle^{a,K}(\beta^{a,K}-\beta^0)\nonumber\\
&&+\sum_K\langle N\rangle^{a,K}(\gamma^{a,K}-\gamma_a^0)=0\label{47}\\
&&\sum_K(\alpha^{a,K}-\alpha_a^0)\langle N\rangle^{a,K}+\sum_K\langle eN\rangle^{a,K}(\beta^{a,K}-\beta^0)\nonumber\\
&&+\sum_K\langle N^2\rangle^{a,K}(\gamma^{a,K}-\gamma_a^0)=0\label{48}
\end{eqnarray}
$\alpha_a^0$ and $\gamma_a^0$ can then be determined from Eqs. (\ref{47})-(\ref{48}), i.e., 
\begin{eqnarray}
&&\alpha_a^0=\sum_K\alpha^{a,K}p^{a,K}+\sum_K\langle e\rangle^{a,K}\beta^{a,K}\nonumber\\
&&+\sum_K\langle N\rangle^{a,K}\gamma^{a,K}-\beta^0\langle e\rangle^a-\gamma_a^0\langle N\rangle^a\\
&& A_{NN}^a\gamma_a^0=\sum_K\alpha^{a,K}\langle N\rangle^{a,K}+\sum_K\langle eN\rangle^{a,K}\beta^{a,K}\nonumber\\
&&+\sum_K\langle N^2\rangle^{a,K}\gamma^{a,K}-A_{eN}^a\beta^0
\end{eqnarray}
where $A_{NN}^a=\langle N^2\rangle^a-\langle N\rangle^a\langle N\rangle^a$ and $A_{eN}^a=\langle eN\rangle^a-\langle e\rangle^a\langle N\rangle^a$ Thus, $\gamma_a^0$ and $\alpha_a^0$ are only a function of $\beta^0$ and system $a$ properties. Furthermore, the evolutions of subsystem properties can be calculated by
\begin{eqnarray}
&&\frac{d\alpha^{a,K}}{dt}=-\frac{1}{\tau}(\alpha^{a,K}-\alpha_a^0)\\
&&\frac{d\gamma^{a,K}}{dt}=-\frac{1}{\tau}(\gamma^{a,K}-\gamma_a^0)\\
&&\frac{d\beta^{a,K}}{dt}=-\frac{1}{\tau}(\beta^{a,K}-\beta^0)
\end{eqnarray}
Thus, the time evolution of $\alpha^{a,K}$, $\gamma^{a,K}$ and $\beta^{a,K}$ can be determined using properties of system $a$ and $\beta^0$ only. The influence of system $b$ is only via $\beta^0$ so that if a different system $b$ can provide the same $\beta$, the time evolution of system $a$ is the same.

To study the linkage between systems $a$ and $b$, the explicit form of $\beta^0=|C_4|/|C_1|$ is given using fluctuations of the extensive properties of energy and particle number, i.e.,
\begin{eqnarray}
\left|C_1\right| &=& \left|\begin{array}{ccc}
A_{NN}^a & 0 & A_{Ne}^a\\
0 & A_{NN}^b & A_{Ne}^b\\
A_{Ne}^a & A_{Ne}^b & A_{ee}
\end{array}\right|\\
\left|C_4\right| &=& \left|\begin{array}{cccc}
A_{Ns}^a & A_{NN}^a & 0 \\
A_{Ns}^b & 0 & A_{NN}^b \\
A_{es} & A_{Ne}^a & A_{Ne}^b 
\end{array}\right|
\end{eqnarray}
where
\begin{equation}
A_{UV}^{a(b)}=\langle UV\rangle^{a(b)}-\langle U\rangle^{a(b)}\langle V\rangle^{a(b)},\,A_{UV}=A_{UV}^{a}+A_{UV}^{b}
\end{equation}
and $A_{UV}^{a(b)}$ is the fluctuation of extensive properties $U$ and $V$ in system $a(b)$, while $A_{UV}$ is the sum of the fluctuations of systems $a$ and $b$.

Now, if system $b$ is much smaller than system $a$, i.e., if $A_{UV}^{a}\gg A_{UV}^{b}$ for any set of extensive properties, $\beta^0\to\tilde{\beta}^a$ where
\begin{equation}
\tilde{\beta}^a\equiv\lim_{\frac{A_{UV}^{b}}{A_{UV}^{a}}\to 0}\frac{\left|C_4\right|}{\left|C_1\right|} =\left|\begin{array}{cc}
A_{es}^a & A_{Ne}^a\\
A_{Ns}^a & A_{NN}^a
\end{array}\right|\left|\begin{array}{cc}
A_{ee}^a & A_{eN}^a \\
A_{eN}^a & A_{NN}^a 
\end{array}\right|
\end{equation} 
The stable equilibrium temperature of system $b$ is $\tilde{\beta}^a$ if the state of system $a$ remains in a nonequilibrium state (for example, if system $a$ relaxes much slower than system $b$ or is controlled by some external interaction). Thus, $\tilde{\beta}^a$ is a temperature measurement of system $a$, which can be used as an expression for the experimental measurement of system $a$ in nonequilibrium. In addition, $\tilde{\beta}^a$ also turns out to be the $\beta^0$ in the equation of motion for a single system relaxation, i.e., when system $b$ is not present.

Now, if the system $b$ is in stable equilibrium and much larger than system $a$, i.e., $A_{UV}^{a}\ll A_{UV}^{b}$ for any set of extensive properties, $\beta^0\to\tilde{\beta}^b=1/k_bT_b$, and $T_b$ is the temperature of system $b$. In this case, system $b$ acts as a heat reservoir at temperature $T_b$.

\section{Equation of motion for interacting systems with heat and mass diffusion}
For the case when both heat and mass diffusion are present, four conservation laws hold: probability conservation for both system $a$ and system $b$, and total energy and total particle number conservation for the composite system. Thus, the generators of motion are $\{\hat{I}_a,\hat{I}_b,\hat{N},\hat{H}\}$ constrained by 
\begin{eqnarray}
I^a&=&\sum_i p_i^a=1\\
I^b&=&\sum_i p_i^b=1\\
N&=&\sum_i N_i^ap_i^a+\sum_i N_i^bp_i^b=\text{const}\\
E&=&\sum_i \epsilon_i^ap_i^a+\sum_i \epsilon_i^bp_i^b=\text{const}
\end{eqnarray}
Similar to the derivation in Appendix B, the equation of motion for the case is
\begin{equation}
\frac{dp_j^a}{dt}=\frac{1}{\tau}\frac{\left|\begin{array}{ccccc}
	-p_j^a\ln \frac{p_j^a}{n_j^a} & p_j^a & 0 & \epsilon_j^ap_j^a & N_j^ap_j^a\\
	\langle s\rangle^a & 1 & 0 &\langle e\rangle^a & \langle N\rangle^a\\
	\langle s\rangle^b & 0 & 1 &\langle e\rangle^b & \langle N\rangle^b\\
	\langle es\rangle & \langle e\rangle^a & \langle e\rangle^b & \langle e^2\rangle & \langle eN\rangle\\
	\langle Ns\rangle & \langle N\rangle^a & \langle N\rangle^b & \langle eN\rangle & \langle N^2\rangle\\
	\end{array}\right|}{\left|\begin{array}{cccc}
	1 & 0 &\langle e\rangle^a & \langle N\rangle^a\\
	0 & 1 &\langle e\rangle^b & \langle N\rangle^b\\
	\langle e\rangle^a & \langle e\rangle^b & \langle e^2\rangle & \langle eN\rangle\\
	\langle N\rangle^a & \langle N\rangle^b & \langle eN\rangle & \langle N^2\rangle\\
	\end{array}\right|}
\end{equation}
The numerator of a ratio of determinants on the right can be expanded to yield
\begin{equation}
\det = -p_j^a\ln \frac{p_j^a}{n_j}|C_1|-p_j^a|C_2^a|-\epsilon_j^ap_j^a|C_3|+N_j^ap_j^a|C_4|
\end{equation}
where $|C_1|$, $|C_2^a|$, $|C_3|$, and $|C_4|$ are the minors of the first line of the determinant. By defining 
\begin{equation}
\frac{|C_2^a|}{|C_1|}=\alpha_a, \quad\frac{|C_3|}{|C_1|}=\beta, \quad\frac{|C_4|}{|C_1|}=-\gamma
\end{equation}
the equation of motion transforms into
\begin{equation}
\frac{dp_j^a}{dt}=\frac{1}{\tau}(-p_j^a\ln \frac{p_j^a}{n_j}-p_j^a\alpha_a-\epsilon_j^ap_j^a\beta-N_j^ap_j^a\gamma)
\end{equation}
Using the row vector $\vec{l}_j^a$ of Eq. (\ref{row}) and defining a new column vector $\vec{\mu}_a^0$ of intensive properties, the equation of motion changes to 
\begin{equation}
\frac{dp_j^a}{dt}=\frac{1}{\tau}(-p_j^a\ln \frac{p_j^a}{n_j}-p_j^a\vec{l_j}^a\cdot\vec{\mu}_a^0)
\end{equation}
where the column vector is defined as
\begin{equation}
\vec{\mu}_a^0=\left[\begin{array}{c}
\alpha_a\\ \beta^0\\ \gamma^0
\end{array}\right]
\end{equation}
Thus, the discussion using the concept of hypoequilibrium state given in Secs. III.B and III.C can be repeated here with the only difference being the definition of $\vec{\mu}_a^0$. Furthermore, the discussion in Sec. III.D is simplified here since only probability conservation holds for system $a$ with the consequence that
\begin{eqnarray}
\frac{dp^a}{dt}&=&\sum_K\frac{dp^{a,K}}{dt}=0
\end{eqnarray}
from which, $\alpha_a$ can be calculated, i.e., 
\begin{eqnarray}
&&\alpha_a=\sum_K\alpha^{a,K}p^{a,K}+\sum_K\langle e\rangle^{a,K}\beta^{a,K}\nonumber\\
&&+\sum_K\langle N\rangle^{a,K}\gamma^{a,K}-\beta^0\langle e\rangle^a-\gamma^0\langle N\rangle^a
\end{eqnarray}
Here, $\alpha_a$ is a function of $\beta^0$, $\gamma^0$, and system $a$ properties. Furthermore, the evolutions of subsystem (i.e., subspace) properties can be determined from
\begin{eqnarray}
&&\frac{d\alpha^{a,K}}{dt}=-\frac{1}{\tau}(\alpha^{a,K}-\alpha_a)\\
&&\frac{d\gamma^{a,K}}{dt}=-\frac{1}{\tau}(\gamma^{a,K}-\gamma^0)\\
&&\frac{d\beta^{a,K}}{dt}=-\frac{1}{\tau}(\beta^{a,K}-\beta^0)
\end{eqnarray}
For this case, the time evolution of $\alpha^{a,K}$, $\gamma^{a,K}$ and $\beta^{a,K}$ are determined using properties of system $a$ and $\beta^0$ and $\gamma^0$ only. The influence of system $b$ is via $\beta^0$ and $\gamma^0$, which relates to the energy and particle number fluxes between the two systems.

To study the linkage between systems $a$ and $b$, the explicit form of $\beta^0=|C_3|/|C_1|$ is given using fluctuations of the extensive properties, i.e.,
\begin{eqnarray}
\left|C_1\right|&=&\left|\begin{array}{cc}
A_{ee} & A_{eN}\\
A_{Ne} & A_{NN}
\end{array}\right|,\,\left|C_3\right|=\left|\begin{array}{cc}
A_{es} & A_{eN}\\
A_{Ns} & A_{NN}
\end{array}\right|,\nonumber\\
\left|C_4\right|&=&\left|\begin{array}{cc}
A_{es} & A_{ee}\\
A_{Ns} & A_{eN}
\end{array}\right|
\end{eqnarray}
The measurements of the intensive properties $\tilde{\beta}^a$ and $\tilde{\gamma}^a$ of system $a$ are given as 
\begin{eqnarray}
\tilde{\beta}^a&\equiv&\lim_{\frac{A_{UV}^{b}}{A_{UV}^{a}}\to 0}\frac{\left|C_3\right|}{\left|C_1\right|} =\left|\begin{array}{cc}
A_{es} & A_{eN}\\
A_{Ns} & A_{NN}
\end{array}\right|/\left|\begin{array}{cc}
A_{ee}^a & A_{eN}^a \\
A_{eN}^a & A_{NN}^a 
\end{array}\right|\quad\\
\tilde{\gamma}^a&\equiv&\lim_{\frac{A_{UV}^{b}}{A_{UV}^{a}}\to 0}\frac{\left|C_4\right|}{\left|C_1\right|} =\left|\begin{array}{cc}
A_{es} & A_{ee}\\
A_{Ns} & A_{eN}
\end{array}\right|/\left|\begin{array}{cc}
A_{ee}^a & A_{eN}^a \\
A_{eN}^a & A_{NN}^a 
\end{array}\right|
\end{eqnarray}
When system $b$ is in stable equilibrium and much larger than system $a$, system $b$ acts as a heat and mass reservoir.

\section{System interacting with multiple systems}

If there are $R$ different kinds of interactions, which system $a$ experiences, the equation of motion changes to 
\begin{eqnarray}
\frac{dp_j^a}{dt}=\sum_{r=1}^{R}\left[\frac{1}{\tau^r}\left(-p_j^a\ln \frac{p_j^a}{n_j^a}-p_j^a\vec{l_j^a}\cdot\vec{\mu}_a^r\right)\right]
\end{eqnarray}
Defining
\begin{eqnarray}
\frac{1}{\tilde{\tau}}=\sum_{r=1}^{R}\frac{1}{\tau^r},\, \frac{\tilde{\vec{\mu}}_a^0}{\tilde{\tau}}=\sum_{r=1}^{R}\frac{\vec{\mu}_a^r}{\tau^r}
\end{eqnarray}
the equation of motion is rewritten as
\begin{eqnarray}
\frac{dp_j^a}{dt}=\frac{1}{\tilde{\tau}}(-p_j^a\ln \frac{p_j^a}{n_j^a}-p_j^a\vec{l_j^a}\cdot\vec{\tilde{\mu}}_a^0)
\end{eqnarray}
which recovers the form of Eq. (\ref{EOM_vec}). Thus, the discussion in Secs. III.B and III.C and in Appendix A still hold. The evolution of hypoequilibrium state and the definition of nonequilibrium intensive properties can be applied to the study a network of nonequilibrium systems with non-quasi-equilibrium interactions.

\section{Conclusions}
This paper provides a thermodynamic investigation of interacting systems undergoing heat and/or mass interactions. In order to apply the SEAQT framework to all kinds of systems, the grand canonical ensemble and the grand partition function are used. The evolutions of intensive and extensive properties as well as the Onsager relations of the relaxation process of non-quasi-equilibrium processes in general are discussed. Both temperature and chemical potential measurements to a system in nonequilibrium is explained from a thermodynamic viewpoint, independent of the microscopic interactions taking place in the measurement. The investigation presented here provides a first-principles explanation for the experimental phenomenological measurement. In addition, both heat and mass reservoirs are defined thermodynamically. Finally, a system interacting with multiple systems is discussed showing how the SEAQT framework and the concepts of hypoequilibrium state and nonequilibrium intensive properties can be applied to studying a network of nonequilibrium system, which in turn permits the study of a macro/mesoscopic system with discrete local systems in nonequilibrium.

\section*{Acknowledgment}
Funding for this research was provided by the US Office of Naval Research under Award No. N00014-11-1-0266.
\appendix
\section{Hypoequilibrium state evolution}
The equation of motions of every energy eigenlevel in the $K$th subspace of system $a$ take the form
\begin{eqnarray}
\frac{dp_j^{a,K}}{dt}=&&\frac{1}{\tau}(-p_j^{a,K}\ln \frac{p_j^{a,K}}{n_j^{a,K}}-p_j^{a,K}\alpha-\epsilon_j^{a,K}p_j^{a,K}\beta\nonumber\\
&&-N_j^{a,K}p_j^{a,K}\gamma)
\end{eqnarray}
Defining
\begin{equation}
\vec{l}_j^{a,K}=\left[\begin{array}{ccc}
1 & \epsilon_j^{a,K} & N_j^{a,K}
\end{array}\right],\,\vec{\mu}^{a,K}=\left[\begin{array}{c}
\alpha^{a,K}\\ \beta^{a,K}\\ \gamma^{a,K}
\end{array}\right],\,\vec{\mu}_a^0=\left[\begin{array}{c}
\alpha\\ \beta\\ \gamma
\end{array}\right]
\end{equation}
the equation of motion is written as
\begin{equation}
\frac{dp_j^{a,K}}{dt}=\frac{1}{\tau}(-p_j^{a,K}\ln \frac{p_j^{a,K}}{n_j^{a,K}}-p_j^{a,K}\vec{l}_j^{a,K}\cdot\vec{\mu}_a^0)
\end{equation}
For the $K$th subspace of system $a$, the probability distribution, grand partition function and $\alpha^{a,K}$ are given by
\begin{eqnarray}
&&p_j^{a,K} = n_j^{a,K}e^{-\alpha^{a,K}}e^{-\epsilon_j^{a,K}\beta^{a,K}}e^{-N_j^{a,K}\gamma^{a,K}}\\
&&\Xi^{a,K}(\beta^{a,K},\gamma^{a,K})=\sum_{i=1}^{\#(\mathcal{H}_a^K)} n_i^{a,K}e^{-\beta^{a,K}\epsilon_i^{a,K}-\gamma^{a,K}N_i^{a,K}}\quad\\
&&\alpha^{a,K}=\ln \Xi^{a,K}(\beta^{a,K},\gamma^{a,K})-\ln p^{a,K}
\end{eqnarray}
The equation of motion then simplifies to
\begin{equation}
\frac{dp_j^{a,K}}{dt}=\frac{p_j^{a,K}}{\tau}\vec{l}_j^{a,K}\cdot(\vec{\mu}^{a,K}-\vec{\mu}_a^0)
\end{equation}
Using the relation
\begin{eqnarray}
\ln \frac{p_j^{a,K}}{n_j^{a,K}}&=&-\mu^{a,K}-\epsilon_j^{a,K}\beta^{a,K}-N_j^{a,K}\gamma^{a,K}\nonumber\\
&=&-\vec{l}_j^{a,K}\cdot\vec{\mu}^{a,K}
\end{eqnarray}
and the fact that the degeneracy $n_j^{a,K}$ is a constant, the equation of motion can also be written as
\begin{eqnarray}
&&-\frac{d}{dt}(\vec{l}_j^{a,K}\cdot\vec{\mu}^{a,K})=\frac{1}{\tau}(\vec{l}_j^{a,K}\cdot\vec{\mu}^{a,K}-\vec{l}_j^{a,K}\cdot\vec{\mu}_a^0)\\
&&\vec{l}_j^{a,K}\cdot(\frac{d\vec{\mu}^{a,K}}{dt}+\frac{1}{\tau}\vec{\mu}^{a,K}-\frac{1}{\tau}\vec{\mu}_a^0)=0\label{Hypo-eq-form}
\end{eqnarray}
For any equation of motion that can reduce to the form of Eq. \ref{Hypo-eq-form} above (e.g., multiple interacting nonequilibrium systems), the system remains in a hypoequilibrium state throughout its evolution provided the system's initial state is a hypoequilibrium state. The solution of this equation is,
\begin{equation}
p_j^{a,K} = n_j^{a,K}e^{-\vec{l}_j^{a,K}\cdot\mu^{a,K}(t)}
\end{equation}
and $\mu^{a,K}(t)$ is found from
\begin{eqnarray}
\frac{d\vec{\mu}^{a,K}(t)}{dt}=-\frac{1}{\tau}(\vec{\mu}^{a,K}(t)-\vec{\mu}_a^0(t))
\end{eqnarray}
which governs the evolutions of the nonequilibrium intensive properties. 

For any three eigenstates, $p_i$, $p_j$, and $p_k$, of system $a$, represented by $\vec{l}_i$, $\vec{l}_j$, and $\vec{l}_k$ where for simplicity the superscripts have been omitted, the following relation is found
\begin{equation}\label{A13}
\ln \frac{p_j}{n_j}=-\vec{l}_j\cdot\vec{K}^{ijk}
\end{equation}
provided $\vec{l}_i$, $\vec{l}_j$, and $\vec{l}_k$ are linearly independent, i.e.,
\begin{equation}
\left|\begin{array}{ccc}
1 & \epsilon_i & N_i\\
1 & \epsilon_j & N_j\\
1 & \epsilon_k & N_k\\
\end{array}\right|\neq 0, \quad \text{or } \left|\begin{array}{cc}
\epsilon_j-\epsilon_i & N_j-N_i\\
\epsilon_k-\epsilon_i & N_k-N_i\\
\end{array}\right|\neq 0
\end{equation}
In Eq. (\ref{A13}), $\vec{K}^{ijk}$ is defined as
\begin{equation}
\vec{K}^{ijk}\equiv\left[\begin{array}{ccc}
1 & \epsilon_i & N_i\\
1 & \epsilon_j & N_j\\
1 & \epsilon_k & N_k\\
\end{array}\right]^{-1}\left[\begin{array}{c}
-\ln\frac{p_i}{n_i}\\
-\ln\frac{p_j}{n_j}\\
-\ln\frac{p_k}{n_k}\end{array}\right]
\end{equation}

The time evolution of these three energy eigenlevels (or eigenstates) obeys the following equations:
\begin{eqnarray}
-\frac{d}{dt}(\vec{l}_j\cdot\vec{K}^{ijk})=\frac{1}{\tau}(\vec{l_j}\cdot\vec{K}^{ijk}-\vec{l_j}\cdot\vec{\mu})\\
\vec{l}_j\cdot(\frac{d\vec{K}^{ijk}}{dt}+\frac{1}{\tau}\vec{K}^{ijk}-\frac{1}{\tau}\vec{\mu})=0
\end{eqnarray}
Because $\vec{l}_i$, $\vec{l}_j$, and $\vec{l}_k$ are linearly independent,
\begin{equation}
\frac{d\vec{K}^{ijk}}{dt}+\frac{1}{\tau}\vec{K}^{ijk}-\frac{1}{\tau}\vec{\mu}=0
\end{equation}
If $\vec{l}_i$, $\vec{l}_j$, and $\vec{l}_k$ are in the same $K$th subspace of system $a$ which is in hypoequilibrium with intensive properties $\vec{\mu}^{a,K}$, the initial condition for the equation of motion of $\vec{K}^{ijk}$ is
\begin{equation}
\vec{K}^{ijk}(t=0)=\vec{\mu}^{a,K}
\end{equation}
Thus, the $\vec{K}^{ijk}$ from any three independent energy eigenlevels in the $K$th subspace of system $a$ follows the same ordinary differential equation, i.e., the same time evolution,
\begin{equation}
\vec{K}^{ijk}(t)=\vec{\mu}^{a,K}(t)
\end{equation}
and, therefore, the system keeps is always in a hypoequilibrium state. 
If no linearly independent triplet of $\vec{l}_i$, $\vec{l}_j$, and $\vec{l}_k$ exists in the subspace, one can set $\gamma=0$ or $\beta=0$ for the case when two linearly independent $\vec{l}_i$ and $\vec{l}_j$ exist in the subspace and set both $\gamma=0$ and $\beta=0$ for the case of a single eigenlevel in the subspace.

\section{Equation of motion}
The energy eigenlevels of system $a$ and $b$ are represented by $\{(n_i^a,\epsilon_i^a,N_i^a)\}$ and $\{(n_j^a,\epsilon_j^a,N_j^a)\}$. The state of the system can be represented by two probability distributions among the energy eigenlevels of systems $a$ and $b$ given by $\{p_i^a,p_j^b,\,i,j=1,2,\cdots\}$. The distance between two states is defined here as the Fisher-Rao metric. Equivalently, the square root of the probability distribution $\{p_i^a,p_j^b,\,i,j=1,2,...\}$ can be used to represent the system state. One can prove that the Fisher-Rao metric of the probability space becomes the Euclidean metric in the space of $\{x_i^a,x_j^b,i,j=1,2,...\}$. The distance between states for both representations is given by
\begin{eqnarray}
&&dl = \frac{1}{2}\sqrt{(\sum_i p_i^a(\frac{d\ln p_i^a}{d\theta})^2+\sum_j p_j^b(\frac{d\ln p_j^b}{d\theta})^2)}d\theta\\
&&dl = \sqrt{(\sum_i x_i^a(\frac{d\ln x_i^a}{d\theta})^2+\sum_j x_j^b(\frac{d\ln x_j^b}{d\theta})^2)}d\theta
\end{eqnarray}
where $dl$ is the distance between $p(\theta+d\theta)$ and $p(\theta)$ or $x(\theta+d\theta)$ and $x(\theta)$, and $\theta$ is a continuous parameter.
A property of the system can be defined as a function of state $\{x_i^a,x_j^b\}$ such that:
\begin{eqnarray}
I^a&=&\sum_i (x_i^a)^2=1\label{B3}\\
I^b&=&\sum_j (x_j^b)^2=1\\
N^a&=&\langle N\rangle^{a}=\sum_i N_i^a(x_i^a)^2=\text{const}\\
N^b&=&\langle N\rangle^{b}=\sum_j N_j^b(x_j^b)^2=\text{const}\\
E&=&\langle e\rangle^{a}+\langle e\rangle^{b}=\sum_i \epsilon_i^a(x_i^a)^2+\sum_j \epsilon_j^b(x_j^b)^2=\text{const}\quad\label{B7}\\
S&=&\langle s\rangle^{a}+\langle s\rangle^{b}=-\sum_i(x_i^a)^2\ln(x_i^a)^2-\sum_j(x_j^b)^2\ln(x_i^a)^2\nonumber\\
\end{eqnarray}
where $\langle \dots\rangle^{a(b)}$ indicates the expectation value in system $a(b)$. For interacting systems with heat diffusion only, there are five conservation laws for the first five properties (Eqs. (\ref{B3})-(\ref{B7})). The von Neumann formula for the entropy is used. For a detailed discussion of why, the reader is referred to \cite{Gyftopoulos1997}.
The gradient of a given property in state space is then expressed by
\begin{eqnarray}
\boldsymbol{g_{I^a}} & = & \sum_{i}\frac{\partial I^a}{\partial x_i^a}\hat{e_i^a}+\sum_{j}\frac{\partial I^a}{\partial x_j^b}\hat{e_j^b} = \sum_{i} 2x_i^a\hat{e_i^a}\\
\boldsymbol{g_{I^b}} & = & \sum_{i}\frac{\partial I^b}{\partial x_i^a}\hat{e_i^a}+\sum_{j}\frac{\partial I^b}{\partial x_j^b}\hat{e_j^b} = \sum_{j} 2x_j^b\hat{e_j^b}\\
\boldsymbol{g_{N^a}} & = & \sum_{i}\frac{\partial N^a}{\partial x_i^a}\hat{e_i^a}+\sum_{j}\frac{\partial N^a}{\partial x_j^b}\hat{e_j^b} = \sum_{i} 2x_i^aN_i^a\hat{e_i^a}\\
\boldsymbol{g_{N^b}} & = & \sum_{i}\frac{\partial N^b}{\partial x_i^a}\hat{e_i^a}+\sum_{j}\frac{\partial N^b}{\partial x_j^b}\hat{e_j^b} = \sum_{j} 2x_j^bN_j^b\hat{e_j^b}\quad\\
\boldsymbol{g_{E}} & = & \sum_{i}\frac{\partial E}{\partial x_i^a}\hat{e_i^a}+\sum_{j}\frac{\partial E}{\partial x_j^b}\hat{e_j^b} \nonumber\\
&= &\sum_{i} 2x_i^a\epsilon_i^a\hat{e_i^a}+\sum_{j} 2x_j^b\epsilon_j^b\hat{e_j^b}\\
\boldsymbol{g_S} & = & \sum_{i}\frac{\partial S}{\partial x_i^a}\hat{e_i^a}+\sum_{j}\frac{\partial S}{\partial x_j^b}\hat{e_j^b}= \sum_{i} [-2x_i^a\nonumber\\
&-&2x_i^a \ln(x_i^a)^2]\hat{e_i^a}+\sum_{j} [-2x_j^b-2x_j^b \ln(x_j^b)^2]\hat{e_j^b}\quad\quad
\end{eqnarray}
where $\hat{e}_{i}^a(b)$ is the unit vector for each dimension.

The principle of SEA upon which the equation of motion is based is defined as the direction at any instant of time along which the system state evolves, which has the largest entropy gradient consistent with the conservation constraints. The resulting equation of motion is then expressed as
\begin{equation}\label{B15}
\frac{d\boldsymbol{x}}{dt} =  \frac{1}{\tau(x)}\boldsymbol{g_S}_{\perp L(\boldsymbol{g_{I^a}},\boldsymbol{g_{I^b}},\boldsymbol{g_{N^a}},\boldsymbol{g_{N^b}},\boldsymbol{g_E})}
\end{equation}
where $\tau$, which is a function of system state, is the relaxation time that describes the speed at which the state evolves in state space in the direction of steepest entropy ascent. $L=L(\boldsymbol{g_{I^a}},\boldsymbol{g_{I^b}},\boldsymbol{g_{N^a}},\boldsymbol{g_{N^b}},\boldsymbol{g_E})$ is the manifold spanned by the first five gradients, and $\boldsymbol{g_S}_{\perp L}$ is the perpendicular component of the gradient of the entropy to the hyper-surface that yields to the five conservation laws. The right hand side of Eq. (\ref{B15}) takes the form of a ratio of Gram determinants. The explicit form of this equation using $\{p_i^a,p_j^b\}$ is given by Eq. (\ref{EOM_heat}).

\bibliography{Grand_20160111}
\end{document}